\documentstyle[aps,prl,preprint,floats,epsfig]{revtex}

\def\Kz    { K^{\circ} }
\def\Kp    { K^{+} }

\def\Bz    { B^{\circ} }
\def\Bp    { B^{+} }

\def\etal  {{\em et al.}}

\begin{document}

\preprint{\tighten\vbox{\hbox{\hfil CLNS 00-1680}
                        \hbox{\hfil CLEO 00-12}
}}


\title{Study of $B$ Decays to Charmonium
States $B\to\eta_c K$ and $B\to\chi_{c0} K$ }

\author{CLEO Collaboration}
\date{\today}

\maketitle
\tighten

\begin{abstract}
In a sample of $9.66\times 10^6 B\bar{B}$ pairs collected
with the CLEO detector we make the first observation of $B$ decays 
to an $\eta_c$ and a kaon. We measure branching fractions
${\cal B}(\Bp\to\eta_c\Kp) = ( 0.69^{+0.26}_{-0.21} \pm 0.08 \pm 0.20 )\times 10^{-3}$
and 
${\cal B}(\Bz\to\eta_c\Kz) = ( 1.09^{+0.55}_{-0.42} \pm 0.12 \pm 0.31 )\times 10^{-3}$,
where the first error is statistical, the second is systematic and the third
is from the $\eta_c$ branching fraction uncertainty.
From these we extract the $\eta_c$ decay constant in the factorization
approximation, $f_{\eta_c} = 335 \pm 75$ MeV.
We also search for $B$ decays to a $\chi_{c0}$
and a kaon. No evidence for a signal is found and we set $90\%$ CL
upper limits:
${\cal B}(\Bp\to\chi_{c0}\Kp) < 4.8\times 10^{-4}$
and 
${\cal B}(\Bz\to\chi_{c0}\Kz) < 5.0\times 10^{-4}$.
\end{abstract}

\newpage

{
\renewcommand{\thefootnote}{\fnsymbol{footnote}}

\begin{center}
K.~W.~Edwards,$^{1}$
R.~Janicek,$^{2}$ P.~M.~Patel,$^{2}$
A.~J.~Sadoff,$^{3}$
R.~Ammar,$^{4}$ A.~Bean,$^{4}$ D.~Besson,$^{4}$ R.~Davis,$^{4}$
N.~Kwak,$^{4}$ X.~Zhao,$^{4}$
% ...
\end{center}
 
\small
\begin{center}
$^{1}${Carleton University, Ottawa, Ontario, Canada K1S 5B6 \\
and the Institute of Particle Physics, Canada}\\
$^{2}${McGill University, Montr\'eal, Qu\'ebec, Canada H3A 2T8 \\
and the Institute of Particle Physics, Canada}\\
$^{3}${Ithaca College, Ithaca, New York 14850}\\
$^{4}${University of Kansas, Lawrence, Kansas 66045}\\
% ...
\end{center}

\setcounter{footnote}{0}
}
\newpage

Two-body $B$ decays to a charmonium state and a kaon have recently received
substantial attention because of their importance for studies
of CP violation in determining the angles of the CKM unitarity triangles
and because of the observation of the unexpectedly large $B$ decay
rate to $\eta^{\prime} X$ \cite{cleo_eta_prime}. Among several theoretical
explanations for the latter, a substantial intrinsic charm component
of the $\eta^{\prime}$ has been proposed \cite{zhitn,ccbar}. 
If this is the case, the $\eta^{\prime}$ can be produced by the axial part
of the $b\to c\bar{c} s(d)$ process, which also produces the $\eta_c$.
Exclusive $B$ decays to charmonium states are also of theoretical interest
as a testing ground for the QCD calculations of quark dynamics
and factorization. In the absence of enhancing mechanisms, the $B$ decay
rate to $\eta_c X$ is expected to be comparable to that for 
the $B$ decay to $J/\psi X$
\cite{deshpandetrampetic,ahmady-mendel,gourdin,hwang-kim,colangelo}.
Experimentally, little is known about $\eta_c$ production in $B$ decays.
The only published result is from a 1995 CLEO study \cite{etci}, 
which used $2.02~fb^{-1}$ of data collected at $\Upsilon(4S)$
and obtained an upper limit on 
inclusive $\eta_c$ production: ${\cal B}(B\rightarrow \eta_c X) < 0.9\% $ 
at $90\%$ CL. 

The color-singlet production of $\chi_{c0}$ in $B$ decays vanishes
in the factorization approximation as a consequence of spin-parity conservation. 
However, the color-octet mechanism \cite{bbyl,charmNLO} allows for 
the production of the $\chi_{c0}$ P-wave $0^{++}$ state via 
the emission of a soft gluon.
No information on $B$ decays to $\chi_{c0}$ 
is available at present \cite{pdg98}.

In this Letter we report results from the analysis of $9.13~fb^{-1}$ of 
$e^+e^-$ annihilation data
collected with the CLEO detector at the Cornell Electron Storage Ring (CESR), 
taken at the $\Upsilon(4S)$ energy,
corresponding to $9.66\times 10^6$ produced $B\bar{B}$ pairs.
In addition, $4.35~fb^{-1}$ of integrated luminosity were taken 
60 MeV below the $\Upsilon(4S)$ resonance in order to study 
backgrounds from light quark production (referred to as continuum).

The data were taken with two configurations of the CLEO
detector, called CLEO II and CLEO II.V.
In the CLEO II configuration of the detector \cite{cleo20},
charged particle tracking is provided by three cylindrical
drift chambers immersed in axial solenoidal magnetic field of 1.5 T.
Charged particle identification (PID) is made possible by 
a time-of-flight system (TOF) outside of the outermost tracking chamber 
and by the measurement of specific ionization loss ($dE/dX$) in 
the tracking system.
Photon and electron identification is provided by
a high resolution electromagnetic CsI (Tl) calorimeter.
The muon system is the outermost subdetector
consisting of three superlayers of wire counters 
interspersed with steel at different absorption lengths.
The CLEO II.V configuration differs from the CLEO II configuration in that 
the innermost drift chamber was replaced by a three layer double-sided
silicon vertex detector \cite{cleo25} and that
a helium-propane gas mixture, instead of argon-ethane, was used 
in the main drift chamber.
These changes led to improved momentum and $dE/dX$ resolution.

We reconstruct the $\eta_c$ in the decay modes
$\eta_c\to\phi\phi\to K^{+}K^{-}K^{+}K^{-}$ and 
$\eta_c\to K^{\circ}_S K^\pm \pi^\mp$
\footnote{Charge conjugation is implied throughout this Letter.}.
The $\chi_{c0}$ is searched for in its decay modes $\chi_{c0}\to K^{+}K^{-}$ 
and $\chi_{c0}\to \pi^{+}\pi^{-}$. 
For calculation of efficiencies, we use the branching fractions 
${\cal B}(\chi_{c0}\to K^{+}K^{-}) = (0.586 \pm 0.086)\%$
and ${\cal B}(\chi_{c0}\to \pi^{+}\pi^{-}) = (0.496 \pm 0.066)\%$ obtained by
averaging the PDG values \cite{pdg98} with the recent BES results \cite{beschiz}.
The detector simulation is based upon GEANT 3 \cite{geant}.

Candidate primary tracks must be well measured and come from the event 
vertex. Neutral kaons are identified as 
a $\pi^{+}\pi^{-}$ pair coming from a displaced vertex. The mass resolution 
is 3.6 MeV and we select events within 8 MeV of the $K^{\circ}_S$ mass \cite{pdg98}.
A phi meson candidate is selected as a $K^+K^-$ pair in the mass window 
$1.00 < M(K^+K^-) < 1.04$ GeV.

We reconstruct the $B$ mesons by combining an $\eta_c$ or a $\chi_{c0}$
with a charged or neutral kaon. The candidate events are identified by using
the difference between the reconstructed and beam energies,
$\Delta E = E(B) - E_{beam}$, and the beam 
constrained mass, $M(B) = \sqrt{ E_{beam}^2 - |\vec{p}_B|^2 }$.
The resolution of $\Delta E$ is about 17 (15) MeV for CLEO II (CLEO II.V).
The uncertainty in $M(B)$ is about 2.6 MeV and is dominated by 
the beam energy spread.
Events for which multiple combinations pass the selection criteria
are assigned a weight equal to the inverse of the number of 
candidates passing the selection. The average number of candidates
per event is about 1.3 for decays in the $\eta_c\to K^{\circ}_S K^\pm \pi^\mp$
submode and is less than 1.1 for all other channels.

To minimize the combinatorial background coming from other $B$ decays
and from continuum production we impose PID criteria.
For the $\eta_c\to\phi\phi$ submode
we require that the charged kaons have $dE/dX$ and TOF measurements
within 3 standard deviations ($\sigma$) of the expected values, 
when such measurements are present. The 
$\eta_c\to K^{\circ}_S K^\pm\pi^\mp$ submode 
has more background and the PID consistency requirements in this case 
are at a more stringent level of $2\sigma$, and
at least one of the PID measurements has to be present for 
the charged kaon candidates.
The secondary tracks from the $\chi_{c0}$ decay 
have momenta between 1.0 and 2.7 GeV, where there is little PID
separation between kaons and pions, therefore, no PID requirements are imposed.

Most of the background comes from continuum production, 
which at CLEO has jet-like characteristics.
We minimize this background by employing the ratio $R2$ of
the second and zeroth Fox-Wolfram moments \cite{fw}.
For isotropic events,
$R2$ is nearly zero and for jet-like events it is close to one. 
We select events with $R2 < 0.25$ for decays with $\eta_c$ 
and $R2 < 0.3$ for decays with $\chi_{c0}$.
In addition, we impose a lepton veto on the bachelor kaon candidate
to remove possible contamination due to $B$ semileptonic decays.

We extract the signal yield by using an unbinned extended 
maximum likelihood fit method described in Ref. \cite{cleolike}.
For the $B\to\eta_c K$ analysis, the fit variables are
the beam constrained mass $M(B)$, the energy difference $\Delta E$,
the $\eta_c$ candidate mass and the cosine of the angle
between the direction of the $B$ candidate
and the beam axis $\cos(\theta_B)$. 
For the $\eta_c\to\phi\phi$ mode we also use the angle $\chi$
between the planes formed by the kaons from the $\phi$ decays
in the $\eta_c$ rest frame \cite{trueman}.

The signal $M(B)$ and $\Delta E$ probability density functions
(PDFs) are parameterized 
by a Gaussian and a sum of two Gaussians with the same mean, respectively.
The $\eta_c$ mass is represented by a Breit-Wigner function
convolved with a sum of two Gaussians with the same mean.
We use $13.2^{+3.8}_{-3.2}$ MeV for the $\eta_c$ width \cite{pdg98}.
The signal shape is expected to vary as $\sin^2(\theta_B)$ and $\sin^2(\chi)$.
The background shape is represented by an end-point function
(the product of a 2nd degree polynomial in $\sqrt{1-(M(B)/E)^2}$ 
and a phase-space factor)
in $M(B)$ and is linear or constant in all other fit variables.
The parameters of the PDF shapes are extracted from the Monte Carlo 
simulation of the signal and backgrounds.
We combine the yields from different data sets and
from different $\eta_c$ submodes by adding the log-likelihood
functions. The yields and efficiencies are given in Table~\ref{effs}.

\begin{table}[tbhp]
\begin{center}
\caption{ Reconstruction efficiencies, including 
${\cal B}(K^{\circ}\to K^{\circ}_S\to\pi^+\pi^-)$,
and signal yields for channels with $\eta_c$.
}
\label{effs} 
\vskip1cm
\begin{tabular}{ lrr }
\hline
Channel  &  Efficiency (\%)  &  Fit yield (events) \\
\hline
$\Bp\to\eta_c\Kp, \eta_c\to\Kz K\pi$ &  13.0  & $18.1_{-5.4}^{+6.2}$ \\
$\Bp\to\eta_c\Kp, \eta_c\to\phi\phi$ &  22.0  & $ 1.4_{-1.0}^{+1.7}$ \\
$\Bz\to\eta_c\Kz, \eta_c\to\Kz K\pi$ &   3.9  & $ 7.5_{-3.2}^{+4.1}$ \\
$\Bz\to\eta_c\Kz, \eta_c\to\phi\phi$ &   6.2  & $ 1.0_{-0.7}^{+1.4}$ \\
\hline
\end{tabular}
\end{center}
\end{table}

Systematic uncertainties from the modeling of the PDF shapes
are included in the fit result by varying the shape parameters
according to their covariance matrices and repeating the fit procedure. 
Systematic errors due to uncertainty 
in the reconstruction efficiency are quoted separately.
An additional source of uncertainty is the $\eta_c$ 
branching fractions;
${\cal B}(\eta_c\to\Kz K\pi)$ has a relative error of 30.9\%
and ${\cal B}(\eta_c\to \phi\phi)$ has a relative error of 39.4\%.
Since the measurements of these branching fractions were made
by experiments running at the $J/\psi$ mass, they
have a common error of 28.3\% due to the uncertainty
of the branching fraction ${\cal B}(J/\psi\to\gamma\eta_c)$.
We quote this common error on the combined result as coming from 
the $\eta_c$ branching fraction uncertainty. 
The remaining errors of 12.4\% and 27.4\% are included 
by smearing the likelihood functions for different sub-modes 
before they are combined.

\begin{figure*}[htb]
\centering
\epsfxsize=165mm
\epsfbox{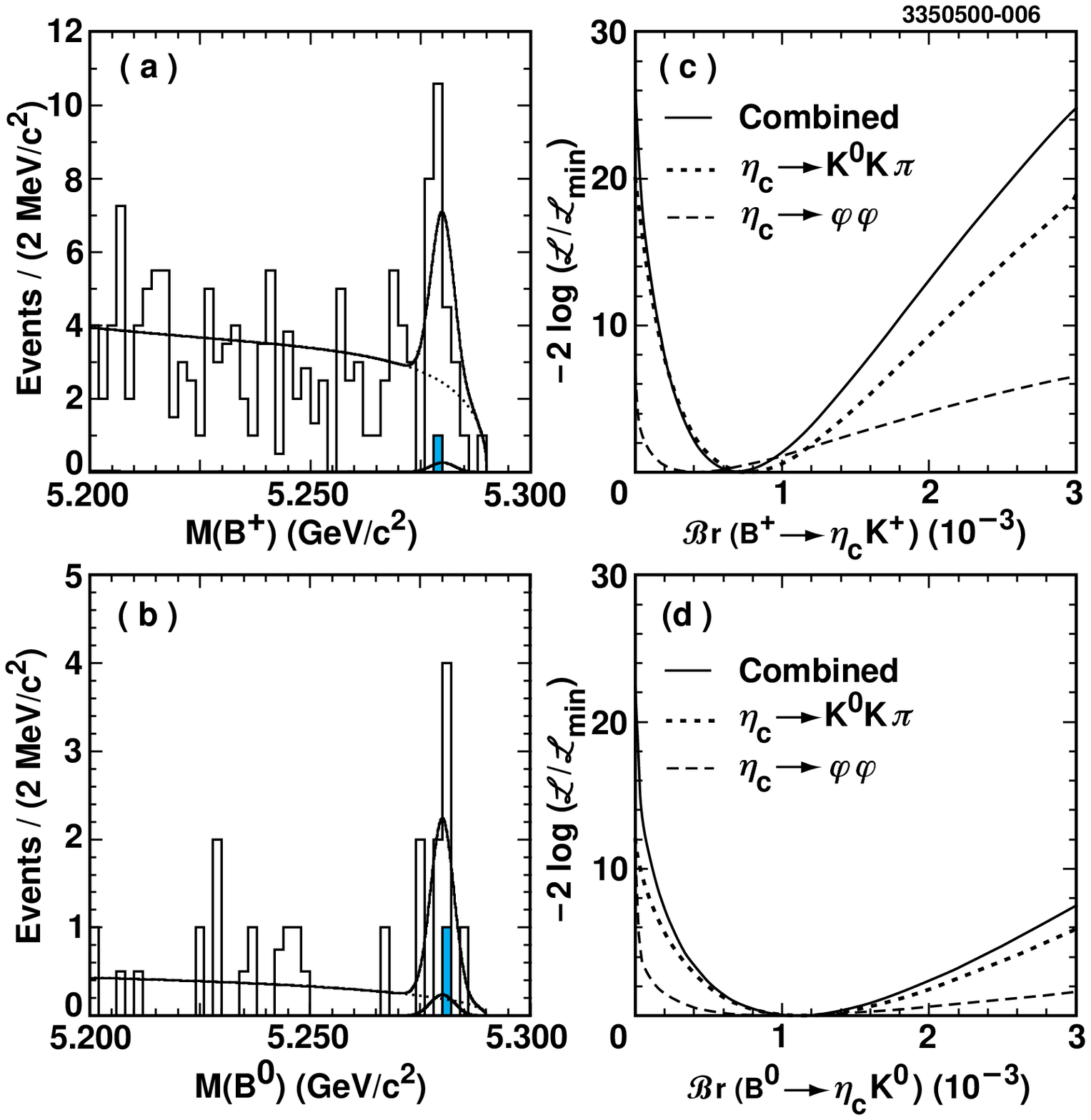}
\caption{  
	Beam constrained mass spectra and the log-likelihood functions 
for charged (a and c) and neutral (b and d) $B$ decay channels.
The mass spectra include both $\eta_c$ channels combined, 
with $\eta_c\to\phi\phi$ also shown separately as the shaded area. 
The solid line in figures (a) and (b) displays
the signal plus background combined shape. The dashed line corresponds
to the background shape only. A cut on the signal likelihood 
using all variables except $M(B)$ is used to make these projections.
}
\label{fig:mblike}
\end{figure*}

Assuming equal production of charged and neutral $B$ mesons
in $\Upsilon(4S)$ decay, the branching fractions are 
${\cal B}(\Bp\to\eta_c\Kp) = ( 0.69^{+0.26}_{-0.21} \pm 0.08 \pm 0.20 )\times 10^{-3}$
and ${\cal B}(\Bz\to\eta_c\Kz) = ( 1.09^{+0.55}_{-0.42} \pm 0.12 \pm 0.31 )\times 10^{-3}$,
where the first error is statistical, the second error is 
from the uncertainty on the reconstruction efficiency
and the third error is due to the uncertainty of ${\cal B}(J/\psi\to\gamma\eta_c)$.
The statistical significances, with PDF shape uncertainties included,
are 5.2 standard deviations for 
the charged $B$ decay and 4.8 for the neutral channel \cite{signif}.
The confidence level of the fits are 54\% and 76\%
for the charged and neutral decay modes, respectively.
The combined $M(B)$ projections and the log-likelihood functions 
of the branching fractions are shown in Fig.~\ref{fig:mblike}.

As a cross-check we have done a counting analysis using more stringent
selection criteria $|M(B)-5.280| < 0.007$ GeV, 
$|\Delta E| < 0.040$ GeV and $|M(\eta_c)-2.9798| < 0.025$ GeV.
The results are statistically consistent with those stated above.
We have checked the sensitivity of our result
to the large spread  of the $\eta_c$
width measurements \cite{pdg98} by repeating the analysis using
values of 10 and 24 MeV for the $\eta_c$ width. The central 
values of the branching fractions are $0.64\times 10^{-3}$
and $0.78\times 10^{-3}$ for the charged
decay and $1.02\times 10^{-3}$ and 
$1.30\times 10^{-3}$ for the neutral decay, 
correspondingly. 

The branching fraction for the decay $B\to\eta_c K$ can be related to that 
of $B\to J/\psi K$ \cite{pdg98} in the factorization approximation by taking
into account the phase space difference and hadronic current dynamics
\cite{deshpandetrampetic,ahmady-mendel,gourdin,hwang-kim,colangelo}.
The ratio of decay rates
is proportional to the ratio of the decay constants squared:
$\Gamma(B\to\eta_c K)/\Gamma(B\to J/\psi K) = D ( f_{\eta_c}/f_{J/\psi} )^2$,
where the coefficient $D$ expresses the evaluation of the decay dynamics.
We calculate the ratio
of the decay constants from the measured branching fractions
using weighted averages of charged and neutral modes and 
the theoretical estimates of $D$.
Our results are consistent with the phenomenological expectations
given in Table~\ref{theorycompar}.
Using the $J/\psi$ decay constant evaluated from dilepton 
rates, $f_{J/\psi} = 405 \pm 14$ MeV \cite{neubertstech},
and predictions of $D$ by Ahmady and Mendel \cite{ahmady-mendel},
we obtain $f_{\eta_c} = 335 \pm 52 \pm 47 \pm 12 \pm 25$ MeV,
where the first error is due to the statistical and systematic errors
on the exclusive branching fractions, the second error is due to the $\eta_c$
branching fractions, the third
error reflects the uncertainty in the $J/\psi$ decay constant,
$f_{J/\psi}$, and the last error is due to $D$.

\begin{table}[tbhp]
\begin{center}
\caption{ Theoretical estimates of $\Gamma(B\to\eta_c K)/\Gamma(B\to J/\psi K)$.
The first, second and fourth columns give phenomenological evaluations
of different quantities. The third column lists our estimate of the decay constant
ratio for a given model, where the first error originates from the branching
fraction uncertainties and the second is due to the quoted error in $D$. }

\label{theorycompar} 
\vskip1cm
\begin{footnotesize}
\begin{tabular}{ l l l l l }
\hline
$D$ &   $f_{\eta_c}/f_{J/\psi}$ & $f_{\eta_c}/f_{J/\psi}$ (exp.)
    & $\frac{\Gamma(B\to\eta_c K)}{\Gamma(B\to J/\psi K)}$ & Ref. \\
\hline
$\cong  2.68$   & $0.78 \pm 0.13$ 
       & $0.54 \pm 0.10 \pm 0.00$
       & $1.64 \pm 0.27$ & \cite{deshpandetrampetic}  \\
$1.12 \pm 0.17$ & $1.20 \pm 0.04$ 
       & $0.83 \pm 0.17 \pm 0.06$
       & $1.6  \pm 0.2 $ & \cite{ahmady-mendel}    \\
$1.11 \pm 0.15$ & $0.99$          
       & $0.84 \pm 0.17 \pm 0.06$
       & $[0.94, 1.24]$  & \cite{gourdin}   \\
$1.11 \pm 0.15$ & $1.03 \pm 0.07$ 
       & $0.84 \pm 0.17 \pm 0.06$
       & $1.14 \pm 0.17$ & \cite{hwang-kim}     \\
$1.43 \pm 0.29$ & $0.81 \pm 0.05$ 
       & $0.73 \pm 0.16 \pm 0.09$
       & $0.94 \pm 0.25$ & \cite{colangelo} \\
\hline
\end{tabular}
\end{footnotesize}
\end{center}
\end{table}

To search for the decay $B\to\chi_{c0}K$ we follow the same
procedure as described above. Part of the background comes 
from $B$ decays to $J/\psi$ or $D$ mesons, such as
$B\to D\pi, D\to K\pi$.
In these cases the reconstructed $B$ momentum is close
to that of the signal and the background peaks in $M(B)$. 
About one half of such background is removed
by vetoing leptons and $K\pi$ combinations in the vicinity of
charged and neutral $D$ mesons masses.
We do not model the behavior of the remaining 
background. Instead, we make a $\pm 7$ MeV cut on $M(B)$.
Only the $\Delta E$ and $M(\chi_{c0})$ variables are used to extract the signal. 
Mistaking pions from $\chi_{c0}$ decay as kaons
shifts the $B$ energy upwards by 120 MeV. 
The $\Delta E$ region is made asymmetric
to keep event samples from different $\chi_{c0}$ decay channels
from overlapping ($-150 < \Delta E < 60$ MeV for the
$\chi_{c0}\to K^{+}K^{-}$ submode and $-60 < \Delta E < 150$ MeV for
the $\chi_{c0}\to \pi^{+}\pi^{-}$ submode).
In spite of the lepton veto some of the background due to
$\psi\to\mu^+\mu^-$ decays remains in the sample because of  
restricted muon system acceptance.
This background contributes to the $M(\chi_{c0})$
sideband of $\chi_{c0}\to K^{+}K^{-}$ submode,
hence we restrict the signal plane to $M(\chi_{c0}) > 3.28$ GeV.
We remove any possible contribution from $\chi_{c2}\to\pi^{+}\pi^{-}$
or $K^{+}K^{-}$ by imposing a skew veto cut $\Delta E > M(\chi_{c0}) -3.5$ GeV.

The final results were extracted from the limited region of 
$\Delta E$ and $M(\chi_{c0})$.
The same unbinned maximum likelihood procedure was used as for 
the $B\to\eta_c K$ analysis.
The observed yields are not statistically significant and the resulting 
90\% confidence level upper limits are $0.48\times 10^{-3}$ 
and $0.50\times 10^{-3}$ for the charged and neutral modes, respectively \cite{deful}.

In summary, we have observed the decay $B\to\eta_c K$ in both charged and
neutral modes with branching fractions similar to those for $B\to J/\psi K$.
By comparing the rates of the decays with $\eta_c$ and $J/\psi$,
we have extracted the $\eta_c$ decay constant. 
The channel $\Bz\to\eta_c\Kz$ can be used to extract the value of 
$\sin( 2 \beta)$ via measurement of time-dependent asymmetry.
We have also set upper limits on $B\to\chi_{c0}K$ decays that
restricts a possible enhancement of the $\chi_{c0}$ production 
due to the color octet mechanism.

We gratefully acknowledge the effort of the CESR staff in providing us with
excellent luminosity and running conditions.
This work was supported by 
the National Science Foundation,
the U.S. Department of Energy,
the Research Corporation,
the Natural Sciences and Engineering Research Council of Canada, 
the A.P. Sloan Foundation, 
the Swiss National Science Foundation, 
the Texas Advanced Research Program,
and the Alexander von Humboldt Stiftung.


\begin{thebibliography}{99}
\bibitem{signif}{
The statistical significance, in terms of the number of standard deviations,
is obtained as $\sqrt{ 2 ( {\cal L}(0) - {\cal L}({\cal B}_{max}) ) }$.
}
\bibitem{deful}{
The 90\% confidence level upper limit on the branching fraction ${\cal B}^{90}$
is defined by integrating the likelihood function:
$\int_0^{{\cal B}^{90}} {\cal L}({\cal B}) d {\cal B} = \int_0^\infty {\cal L}({\cal B}) d {\cal B}$.
}
\end{thebibliography}
\end{document}